\documentclass[12pt,a4paper]{article}
\usepackage[T1]{fontenc}
\usepackage[margin=28mm]{geometry}
\usepackage{amsmath,amssymb,amsthm,mathtools,bm}
\usepackage{booktabs,tabularx,array}
\usepackage{enumitem,etoolbox,needspace,placeins}
\usepackage[numbers,sort&compress]{natbib}
\usepackage{xurl}
\usepackage{hyperref}
\usepackage{setspace}
\hypersetup{hidelinks,pdftitle={Feedback-Aware Tuning of Recursive Q-Learning},pdfauthor={Masahiro Kojima}}
\AtBeginEnvironment{table}{\normalsize\doublespacing}
\AtBeginEnvironment{proposition}{\Needspace{6\baselineskip}}
\AtBeginEnvironment{theorem}{\Needspace{6\baselineskip}}
\AtBeginEnvironment{assumption}{\Needspace{5\baselineskip}}
\AtBeginEnvironment{corollary}{\Needspace{9\baselineskip}}
\renewenvironment{abstract}{\begin{center}\bfseries Abstract\end{center}\noindent\ignorespaces}{\par}
\setlist[enumerate,1]{label=(\alph*),leftmargin=2.2em,itemsep=0.3em}
\newtheorem{theorem}{Theorem}
\newtheorem{proposition}{Proposition}
\newtheorem{corollary}{Corollary}

\newtheorem{assumption}{Assumption}
\theoremstyle{remark}

\newcommand{\R}{\mathbb R}
\newcommand{\E}{\mathbb E}

\newcommand{\cA}{\mathcal A}
\newcommand{\cG}{\mathcal G}
\newcommand{\cP}{\mathcal P}
\newcommand{\cR}{\mathcal R}
\newcommand{\cS}{\mathcal S}
\newcommand{\cT}{\mathcal T}
\newcommand{\cW}{\mathcal W}
\newcommand{\ip}[2]{\langle #1,#2\rangle}
\newcommand{\norm}[1]{\lVert #1\rVert}

\DeclareMathOperator{\tr}{tr}
\DeclareMathOperator{\KL}{KL}

\title{Feedback-Aware Tuning of Recursive Q-Learning}
\author{Masahiro Kojima\thanks{Department of Data Science for Business Innovation, Chuo University, 1-13-27 Kasuga, Bunkyo-ku, Tokyo 112-8551, Japan. E-mail: mkojima263@g.chuo-u.ac.jp.}}
\date{}
\begin{document}
\maketitle

\begin{abstract}
Model choice in backward Q-learning is recursive because a later-stage choice changes the response supplied to an earlier regression and can alter its model-comparison statistic. Separate stagewise criteria do not directly assess the target-stage prediction risk of a completed Q-learning fit. We address this mismatch by treating the entire backward-fitting rule as the unit of comparison and propose feedback-aware soft tuning for sequential multiple assignment randomised trials. Each backward fit uses its own generated responses and is assessed at a common prediction target. The risk criterion retains downstream effects on upstream comparisons, while a separate correction accounts for estimating the final exponential weights from the same observations. For a fixed finite library of smooth recursive maps, we establish an exact risk identity under a Gaussian shift model and an oracle inequality with an explicit adaptation remainder. Under coordinate-representation and moment conditions, these guarantees transfer to prediction risk at any prespecified stage, with the number of stages fixed as sample size increases. A two-stage construction supplies an explicit observable implementation. Numerical studies examine risk estimation and finite-sample performance, and a simulated attention-deficit/hyperactivity-disorder trial illustrates the relation between comparison feedback and treatment recommendations.
\end{abstract}

\section{Introduction}\label{sec:introduction}
Sequential multiple assignment randomised trials (SMARTs) support the development of dynamic treatment regimes through sequential randomisation \cite{Murphy2005Design,NahumShani2012}. Q-learning estimates these regimes by backward regression, incorporating fitted later-stage optimal values into earlier-stage pseudo-outcomes \cite{Murphy2003,Murphy2005,Schulte2014}. This backward construction makes model choice recursive. Changing a downstream selection or averaging rule changes the response used upstream and can alter the next model-comparison statistic, even when the upstream candidate regression models and comparison criterion are unchanged. Consequently, stagewise comparisons need not rank complete Q-learning specifications by their initial-stage Q-function prediction risk. The central difficulty is therefore not only that estimation error propagates backwards, but that a downstream data-dependent fitting decision can change the statistical comparison subsequently performed upstream.

Work on model uncertainty, sequential estimation and risk assessment provides relevant foundations. Frequentist model averaging develops risk-based weights and comparisons with selection \cite{Hansen2007,HjortClaeskens2003,PengYang2022}. For dynamic treatment regimes, \citet{Huang2015} study accumulated misspecification bias and its reduction in backward Q-learning, while penalised Q-learning, interactive model building and regression-based variable selection address complementary modelling and inferential questions \cite{SongEtAl2015,LaberLinnStefanski2014,BianEtAl2023}. Related methods include information criteria for blip-model selection in G-estimation \cite{WallaceMoodieStephens2019}, treatment-effect-oriented model selection \cite{RollingYang2014}, decision-oriented Bayesian model combination \cite{TallmanWest2024}, and finite-time bounds for fitted value iteration \cite{MunosSzepesvari2008}. Stein risk estimation and aggregation assess data-dependent estimators and their combinations \cite{Stein1981,LeungBarron2006,DalalyanSalmon2012}; work on tuning optimism \cite{TibshiraniRosset2019} includes corrected bounds for linear estimators tuned by Stein's unbiased risk estimate (SURE) \cite{CauchoisAliDuchi2021}. These developments do not directly address how a downstream fit changes the pseudo-outcome and comparison statistic used upstream. We quantify this comparison feedback and incorporate it into observable risk assessment for comparing and combining completed Q-learning fits at a common prediction target.

We propose feedback-aware (FA) soft tuning of Q-learning fits from prespecified regression models and weighting rules across stages. Each specification generates its own backward responses. Exact error identities and a residualised-coefficient representation quantify changes in upstream comparison. A target-stage Stein criterion \cite{Stein1981} differentiates through the complete recursion, rather than holding generated responses fixed. Exponential weights combine completed predictions without refitting; a separate derivative correction accounts for estimating these weights from the same data. For fixed finite smooth libraries, we establish an exact Gaussian risk identity and an oracle inequality with an explicit adaptation remainder. A general first-order theorem transfers these results to full Q-function prediction risk at any prespecified stage of a $K$-stage SMART, with $K$ fixed as sample size increases. A fully observed two-stage model verifies the conditions using observable regression coordinates and estimated transport coefficients. Numerical studies compare FA with recursive Akaike weighting, and a simulated attention-deficit/hyperactivity-disorder trial illustrates comparison feedback and its relation to treatment recommendations.

Section~\ref{sec:feedback} establishes the recursive comparison mechanism. Section~\ref{sec:fa-tuning} defines the combined estimate, its risk-based weights and its Gaussian risk guarantee. Section~\ref{sec:implementation} transfers the guarantee to SMART risk and constructs an observable two-stage implementation. Section~\ref{sec:numerical} evaluates risk estimation and finite-sample performance and presents a simulated-trial diagnostic. Section~\ref{sec:discussion} concludes. Proofs and computational details are in the Supplementary Material.

\section{Why model comparison is recursive}
\label{sec:feedback}

Why are separate stagewise criteria insufficient for comparing complete Q-learning fits? We first decompose target-stage prediction risk, then show how downstream fitting changes the statistic used in an upstream nested-model comparison.

\subsection{Q-learning fits and the prediction target}
For participant $i=1,\ldots,n$ and stage $s=1,\ldots,K$, let $H_{si}$ be the history before treatment $A_{si}\in\cA_s(H_{si})$, and let $Y_{si}$ be the reward before the next decision. Each available action set is finite. Histories are nested, with $(H_s,A_s,Y_s)$ included in $H_{s+1}$. We assume independent participant trajectories, consistency, no interference, sequential randomisation and positivity. For a terminal outcome, set $Y_s=0$ before the last stage.

Write $V_s^*(h)=\max_a Q_s^*(h,a)$ and $V_{K+1}^*=0$. The optimal Q-function is
\begin{equation}
 Q_s^*(h,a)=\E\{Y_s+V_{s+1}^*(H_{s+1})\mid H_s=h,A_s=a\}.
 \label{eq:true-q}
\end{equation}
A prespecified tie-breaking rule designates a maximiser $d_s^*(h)$ when several actions attain the same maximum. A Q-learning specification fixes the regression models and the selection or weighting rule at each stage. Under specification $r$, the backward fit uses
\begin{equation}
 \widetilde Y_s^r=Y_s+\widehat V_{s+1}^r(H_{s+1}),\qquad
 \widehat Q_s^r=\sum_{m=1}^{M_s}\widehat w_{s,m}^r\widehat Q_{s,m}^r,
 \quad \widehat V_s^r(h)=\max_a\widehat Q_s^r(h,a).
 \label{eq:recursion}
\end{equation}
The recursion starts with $\widehat V_{K+1}^r=0$. The weights belong to the simplex $\cW_M=\{a\in[0,1]^M:\sum_m a_m=1\}$; hard selection uses a vertex. A stagewise regression model can have different fitted coefficients under different Q-learning specifications because the supplied response changes. Below, $m$ indexes a regression model within a stage, whereas $j$ indexes a specification of the full backward fit. Model selection and model averaging describe choices within a stage; the resulting Q-learning estimate also depends on how those choices are combined across stages.

For a fixed evaluation law $\nu_t$ on target-stage histories and treatments, define $\ip{f}{g}_t=\E_{\nu_t}(fg)$ and $\norm{f}_t^2=\ip{f}{f}_t$. The prediction risk is $\cR_{t,n}^r=\E\norm{\widehat Q_t^r-Q_t^*}_t^2$, where the outer expectation is taken over the data used for the backward fit. The evaluation draw under $\nu_t$ is independent of these data, and $\nu_t$ is common to all Q-learning specifications. In our two-stage implementation it is the SMART population law of $(H_1,A_1)$. This is Q-function prediction risk, not the value of a fitted treatment regime. In local asymptotic statements the true functions are indexed by $n$; elsewhere that index is suppressed.

\subsection{How prediction errors propagate across stages}
Let $(\cP_s\psi)(h,a)=\E\{\psi(H_{s+1})\mid H_s=h,A_s=a\}$ and let $\cT_s\psi$ add the conditional mean reward. A fitted $\psi$ is held fixed when this population expectation is taken. Put $e_s^r=\widehat Q_s^r-Q_s^*$ and $\zeta_s^r=\widehat Q_s^r-\cT_s\widehat V_{s+1}^r$. The latter is fitting error relative to the population target generated by the realised downstream fit.

Set $\Gamma_s(h,a)=V_s^*(h)-Q_s^*(h,a)$ and $(\cS_sF)(h)=F\{h,d_s^*(h)\}$. The exact maximisation remainder is
\begin{equation}
 \rho_s(F)(h)=\max_{a\in\cA_s(h)}
 \left[F(h,a)-F\{h,d_s^*(h)\}-\Gamma_s(h,a)\right].
 \label{eq:switching}
\end{equation}
It is nonnegative. Let $\cG_s=\cP_s\cS_{s+1}$ and let an empty product of these operators be the identity. The next result separates transported fitting errors from maximisation remainders and expands their joint risk, showing why stagewise risks cannot generally be assessed separately.

\begin{proposition}
For each realised recursive fit, $e_K^r=\zeta_K^r$ and $e_s^r=\zeta_s^r+\cG_se_{s+1}^r+\cP_s\rho_{s+1}(e_{s+1}^r)$ for $s<K$. Consequently,
\begin{equation}
 e_t^r=\sum_{s=t}^K\cG_{t:s-1}\zeta_s^r+
       \sum_{s=t+1}^K\cG_{t:s-2}\cP_{s-1}\rho_s(e_s^r),
 \qquad \cG_{j:k}=\cG_j\cdots\cG_k.
 \label{eq:unrolled}
\end{equation}
If $L_{t,s}^r=\cG_{t:s-1}\zeta_s^r$, $L_t^r=\sum_{s=t}^KL_{t,s}^r$, and $N_t^r$ is the second sum, then, whenever the terms are integrable,
\begin{equation}
 \cR_{t,n}^r=\sum_{s,u=t}^K\E\ip{L_{t,s}^r}{L_{t,u}^r}_t
            +2\E\ip{L_t^r}{N_t^r}_t+\E\norm{N_t^r}_t^2.
 \label{eq:exact-risk}
\end{equation}
\end{proposition}

The proof is given in Supplementary Section S6.1.

The cross-stage inner products can have either sign, so negligible treatment switching alone does not make risk additive. For any two Q-learning specifications, subtracting their respective terms $\cS_{s+1}e_{s+1}^r+\rho_{s+1}(e_{s+1}^r)$ gives the difference between their supplied responses, evaluated at $H_{s+1}$. Thus downstream fitting changes the response on which an upstream criterion is computed, not merely a risk contribution to be added after fitting.

\subsection{How downstream fitting changes an upstream comparison}
Proposition 1 identifies the response perturbation entering the preceding regression. We now determine how that perturbation changes the comparison between nested least-squares models. Let $X_s$ contain the smaller (narrow) model's regressors and $Z_s$ the $d_s$ added regressors of the larger (wide) model. Set $W_s=[X_s,Z_s]$ and $M_{X_s}=I-P_{X_s}$, where $P_{X_s}$ projects onto the column space of $X_s$. When the Gram matrices are invertible, subtracting the ordinary least-squares normal equations gives the full coefficient shift, while partialling out $X_s$ gives the added-block shift,
\begin{align}
 \widehat{\bm\theta}_s^r-\widehat{\bm\theta}_s^\circ
 &=(W_s^{\mathsf T} W_s)^{-1}W_s^{\mathsf T}\bm u_s^r,\notag\\*
 \widehat{\bm\gamma}_s^r-\widehat{\bm\gamma}_s^\circ
 &=(Z_s^{\mathsf T} M_{X_s}Z_s)^{-1}Z_s^{\mathsf T} M_{X_s}\bm u_s^r.
 \label{eq:coefficient-shift}
\end{align}
Here $\widehat{\bm\theta}$ is the full wide-model coefficient vector and $\widehat{\bm\gamma}$ its added-block subvector. The superscript $\circ$ denotes the fit to the oracle response $Y_s+V_{s+1}^*(H_{s+1})$, which uses the true continuation value. The vector $\bm u_s^r$ is the response generated under specification $r$ minus this oracle response. These two identities hold for any pair of response vectors fitted on the same design matrices. When both responses come from fitted Q-functions, the coefficient shifts are observable and require no oracle fit.

For a first-order expansion of the standardised added-block coefficient vector, define
$\widehat G_{\bar z,s}=Z_s^{\mathsf T} M_{X_s}Z_s/n$ and
$\bm\eta_{s,n}^r=\sqrt n\,\widehat G_{\bar z,s}^{1/2}
\widehat{\bm\gamma}_s^r/\widehat\sigma_s$, using a common consistent
reference scale $\widehat\sigma_s$ for the recursive and oracle coefficient vectors.
The residual-scale estimates used in the corresponding working-likelihood
fits need not be identical in finite samples; it is sufficient that both
converge in probability to the same population scale $\sigma_s$. Proposition 2 identifies the first-order change in this standardised coefficient vector and hence in the likelihood-ratio statistic used for model comparison.

\begin{proposition}
Let $\bar{\bm z}_s$ be the population residual of the added regressors on the narrow span, $G_{\bar z,s}=\E(\bar{\bm z}_s\bar{\bm z}_s^{\mathsf T})\succ0$, and let $\sigma_s^2>0$ be the residual variance of the limiting homoscedastic oracle model. Suppose the perturbation of the generated response is $n^{-1/2}\bm b_{s+1}(H_{s+1})^{\mathsf T}\bm Z_{s+1,n}^r+r_{s,n}^r$, where $\bm Z_{s+1,n}^r=O_p(1)$, the relevant empirical cross-moment converges to its population counterpart, and $n^{-1/2}\sum_i\bar{\bm z}_{si}r_{si,n}^r=o_p(1)$. Assume also that $\widehat G_{\bar z,s}\to_p G_{\bar z,s}$, $\widehat\sigma_s\to_p\sigma_s$, and replacing population residuals by sample residuals in these normalised moments is negligible. Then the standardised added-block coefficient vector satisfies
\begin{equation}
 \bm\eta_{s,n}^r=\bm\eta_{s,n}^{\circ}+C_s\bm Z_{s+1,n}^r+o_p(1),
 \qquad C_s=\sigma_s^{-1}G_{\bar z,s}^{-1/2}
       \E\{\bar{\bm z}_s\bm b_{s+1}^{\mathsf T}\}.
 \label{eq:feedback-score}
\end{equation}
If $(\bm\eta_{s,n}^{\circ},\bm Z_{s+1,n}^r)\Rightarrow(\bm\eta_s^\circ,\bm Z_{s+1}^r)$ and the wide-model residual mean square converges to $\sigma_s^2$, the corresponding twice-log-likelihood ratio converges to $\|\bm\eta_s^\circ+C_s\bm Z_{s+1}^r\|^2$.
\end{proposition}
The proof is given in Supplementary Section S6.3. The response expansion describes a local perturbation of the continuation value, not an independence assumption. For finite-dimensional linear Q-models, evaluating the coefficient error at the designated optimal action gives the leading term, while the remainder contains treatment switching and any higher-order approximation error. The projected-remainder condition requires these remaining terms to be negligible for the upstream standardised coefficients. Gram and cross-moment convergence and residual-scale consistency can be checked from the regression moments; the response expansion and switching control require verification for the specified trial model. Supplementary Sections S4.7 and S4.10 provide these checks for the two-stage construction.

Thus $C_s$ maps downstream perturbations into the upstream standardised coefficient vector. For unknown variance, write $\mathrm{RSS}_{s,j}$ for the residual sum of squares in model $j$. The twice-log-likelihood ratio statistic is $\Lambda_{s,n}=n\log(\mathrm{RSS}_{s,0}/\mathrm{RSS}_{s,1})$. The Akaike information criterion (AIC) selects the wide model when $\Lambda_{s,n}>2d_s$, whereas its normalised Akaike weight is $\{1+\exp(d_s-\Lambda_{s,n}/2)\}^{-1}$ \cite{Akaike1974}. Equation~\eqref{eq:feedback-score} shows how downstream fitting can change the input to either rule when $C_s\ne0$. Independent fitting samples can remove correlation between the underlying regression scores but not this population response channel. Conversely, $C_s=0$ does not imply finite-sample invariance because the residual scale can change. It removes this particular first-order coefficient-perturbation channel; it is not equivalent to common fixed upstream regressions or to independence of recursive fitting errors.

Together, the two propositions identify what a target-stage comparison must retain: dependence within each backward fit and the interaction of its propagated errors. We therefore keep each Q-learning specification intact and compare the completed Q-function predictions, as developed next.

\section{Feedback-aware soft tuning}\label{sec:fa-tuning}
FA soft tuning combines completed Q-learning fits. We define its output, construct the common-target risk criterion for choosing weights, and account for their estimation when assessing the combined predictor.

\subsection{The combined Q-function estimate}
A \emph{Q-learning specification} fixes the models and selection or weighting rules across stages. As above, $m$ indexes a stagewise model and $j=1,\ldots,J$ a complete specification. Different weighting rules can generate different pseudo-outcomes and upstream fits even with identical model libraries.

For a prespecified target stage $t$, fit each specification to the same $n$ participants, from stage $K$ back to $t$. At every step with $s>t$, maximising that fit's Q-function supplies the continuation value for its preceding regression. Denote the resulting target-stage estimate by $\widehat Q_{t,n}^{(j)}$. FA soft tuning returns
\begin{equation}
 \widehat Q_{t,n}^{\mathrm{FA}}
       =\sum_{j=1}^J\widehat\alpha_{t,j,n}\widehat Q_{t,n}^{(j)}.
 \label{eq:general-output}
\end{equation}
The nonnegative weights sum to one and are chosen for prediction at stage $t$, not separately for each evaluation pair. No regression is refitted after combination; stagewise model averaging remains inside each backward fit. We next construct the risk estimates determining $\widehat\alpha_{t,j,n}$.

\subsection{A risk criterion that retains comparison feedback}
To assess the completed fits on a common local scale, consider the Gaussian shift model $\bm w\sim N_p(\bm\delta,I_p)$. Here $\bm w$ is the observation vector and $\bm\delta$ its mean. The fixed maps $f_j:\R^p\to\R^m$ represent the prediction coordinates of the backward fits, all evaluated at the same $\bm w$. A fixed, known matrix $\bm A\in\R^{m\times p}$ maps $\bm\delta$ to the linear part $\bm A\bm\delta$ of the target; a common intercept is omitted. This model describes local prediction risk, not the distribution of the original outcomes. Section~\ref{sec:implementation} constructs observable SMART coordinates and estimates the target-matrix entries without supplying $Q^*$ or $\bm\delta$ to the fitter.

For a vector-valued map $h$ with input $\bm x$, write $\partial_{\bm x^{\mathsf T}}h=\partial h/\partial\bm x^{\mathsf T}$ for its output-by-input Jacobian. For a scalar function $u$, write $\partial_{\bm x}u=\partial u/\partial\bm x$ for its column vector of partial derivatives. Thus $\partial_{\bm w^{\mathsf T}}f_j(\bm w)$ is $m\times p$, whereas $\partial_{\bm w}S_j(\bm w)$ below is $p$-dimensional. Differentiation with respect to $\bm w$ retains the entire backward recursion while holding the specification and target matrix fixed.

Under Stein's differentiability and integrability conditions, the risk estimate for the map $f_j$ is
\begin{equation}
 S_j(\bm w)
 =\|f_j(\bm w)-\bm A\bm w\|^2
       +2\tr\{\bm A^{\mathsf T} \partial_{\bm w^{\mathsf T}}f_j(\bm w)\}-\|\bm A\|_F^2
 \label{eq:fa-fixed-score}
\end{equation}
with expectation $R_j(\bm\delta)=\E_{\bm\delta}\|f_j(\bm w)-\bm A\bm\delta\|^2$ \cite{Stein1981}. Vector norms are Euclidean and $\|\bm A\|_F$ is the Frobenius norm. The theory uses smooth recursive maps, including smooth Akaike-type weighting and the all-wide linear reference, but excludes discontinuous hard selectors and threshold-shrinkage maps.

The role of the full derivative is explicit in a two-stage map. Split $\bm w=(\bm w_1^{\mathsf T},\bm w_2^{\mathsf T})^{\mathsf T}$ into blocks of dimensions $d_1,d_2$, and let $g_{1,j},g_{2,j}$ be smooth stagewise maps. For $C\in\R^{d_1\times d_2}$ and $D\in\R^{q\times d_2}$, set
\begin{equation}
 f_j(\bm w)=
 \begin{pmatrix}
  g_{1,j}\{\bm w_1+Cg_{2,j}(\bm w_2)\}\\
  Dg_{2,j}(\bm w_2)
 \end{pmatrix},
 \qquad
 \bm A=
 \begin{pmatrix}
  I_{d_1}&C\\
  0&D
 \end{pmatrix}.
 \label{eq:complete-map}
\end{equation}
The first block retains the upstream comparison induced by the downstream fit: $g_{2,j}(\bm w_2)$ changes the input to $g_{1,j}$. The second retains prediction directions transported into the upstream narrow span. Omitting it generally evaluates only part of the target-stage prediction error.

Let $\bm q_j=\bm w_1+Cg_{2,j}(\bm w_2)$ and write $J_1=\partial g_{1,j}(\bm q_j)/\partial\bm q_j^{\mathsf T}$ and $J_2=\partial g_{2,j}(\bm w_2)/\partial\bm w_2^{\mathsf T}$, with each derivative taken with respect to its displayed input. Then
\begin{equation}
 \tr(\bm A^{\mathsf T} \partial_{\bm w^{\mathsf T}}f_j)
 =\tr(J_1)+\tr(C^{\mathsf T} J_1CJ_2)+\tr(D^{\mathsf T} DJ_2).
 \label{eq:feedback-trace}
\end{equation}
The middle term is the comparison-feedback contribution missing from criteria that hold continuation fits fixed. It remains even with independent Gaussian blocks. Thus \eqref{eq:fa-fixed-score} evaluates each complete recursive map against the common prediction target while retaining the dependence identified in Section~\ref{sec:feedback}. These risk estimates now provide a basis for choosing the outer weights.

\subsection{Risk-based weights and the risk of the combined predictor}
Fix a tuning parameter $T>0$ and positive prior weights $\pi_j$ summing to one before observing the data. In the Gaussian model, define
\begin{equation}
 \alpha_j(\bm w)=\frac{\pi_j\exp\{-S_j(\bm w)/T\}}
                   {\sum_k\pi_k\exp\{-S_k(\bm w)/T\}},\qquad
 \bar f(\bm w)=\sum_j\alpha_j(\bm w)f_j(\bm w).
 \label{eq:fa-weights}
\end{equation}
The weight vector $\alpha$ minimises $\sum_j a_jS_j+T\KL(a\Vert\pi)$ over $\cW_J=\{a\in[0,1]^J:\sum_{j=1}^J a_j=1\}$, where $\KL(a\Vert\pi)=\sum_j a_j\log(a_j/\pi_j)$ is the Kullback--Leibler divergence, with $0\log(0/\pi_j)=0$. Smaller $T$ concentrates weight more strongly on fits with smaller estimated risk. It is prespecified, not chosen by further same-data minimisation. Multiplying all risk estimates by a positive constant requires multiplying $T$ by that constant to preserve the weights.

The weights define the combined predictor, but averaging the individual risk estimates does not generally estimate its risk without bias. Averaging predictions changes squared loss, and estimating the weights adds data dependence. To account for both effects, set $\bar{\bm s}=\sum_j\alpha_j\partial_{\bm w}S_j$ and define
\begin{align}
 \mathrm{Disp}_f&=\sum_j\alpha_j\|f_j-\bar f\|^2,\qquad
 \mathrm{Disp}_{\partial S}=\sum_j\alpha_j\|\bm A(\partial_{\bm w}S_j-\bar{\bm s})\|^2,\notag\\
 B&=-\frac2T\sum_j\alpha_j(f_j-\bar f)^{\mathsf T}\bm A\partial_{\bm w}S_j.
 \label{eq:fa-adaptation}
\end{align}
Here $\mathrm{Disp}_f$ measures the spread of fitted predictions, and $\mathrm{Disp}_{\partial S}$ measures the spread of their risk-criterion derivatives after transformation by $\bm A$. Differentiating the outer weights with respect to $\bm w$ gives the signed correction $B$, distinct from feedback within each backward fit. Since $\sum_j\alpha_j(f_j-\bar f)=0$, $B$ is unchanged when $\partial_{\bm w}S_j$ is replaced by $\partial_{\bm w}S_j-\bar{\bm s}$; it measures the weighted association between the two sets of deviations.

The next theorem completes the Gaussian risk assessment. It corrects the individual risk estimates for averaging and weight estimation, and compares the resulting aggregate risk with the risks of the separate fits.

\begin{theorem}
Let $\bm w\sim N_p(\bm\delta,I_p)$, with a fixed, known target matrix $\bm A$, a fixed tuning parameter $T>0$ and fixed positive prior weights. Suppose the finite library consists of $C^2$ maps and, for the mean under consideration,
\begin{equation}
 \E_{\bm\delta}\sum_{j=1}^J
 \left\{\|f_j(\bm w)\|^2+\|\partial_{\bm w^{\mathsf T}}f_j(\bm w)\|_F+\|\partial_{\bm w}S_j(\bm w)\|^2\right\}<\infty.
 \label{eq:gaussian-regularity}
\end{equation}
Then the risk estimate
\begin{equation}
 S_{\mathrm{FA}}(\bm w)=\sum_j\alpha_jS_j-\mathrm{Disp}_f+B
 \label{eq:fa-adapted-score}
\end{equation}
is unbiased for $R_{\mathrm{FA}}(\bm\delta)=\E_{\bm\delta}\|\bar f(\bm w)-\bm A\bm\delta\|^2$. Moreover, $\E_{\bm\delta} \mathrm{Disp}_{\partial S}<\infty$ and
\begin{align}
 R_{\mathrm{FA}}(\bm\delta)
 &\leq\inf_{q\in\cW_J}
       \left\{\sum_jq_jR_j(\bm\delta)+T\KL(q\Vert\pi)\right\}
       +\frac{\E_{\bm\delta} \mathrm{Disp}_{\partial S}}{T^2}\notag\\
 &\leq\min_j\{R_j(\bm\delta)+T\log(1/\pi_j)\}
       +\frac{\E_{\bm\delta} \mathrm{Disp}_{\partial S}}{T^2}.
 \label{eq:fa-oracle}
\end{align}
If $\sup_{j,\bm w}\|\bm A\partial_{\bm w}S_j(\bm w)\|\leq L$, the remainder is at most $L^2/T^2$ for every fixed Gaussian mean $\bm\delta$.
\end{theorem}

The risk-identity and oracle-inequality proofs are given in Supplementary Sections S1.3 and S1.4, respectively; Section S1.1 verifies the integration-by-parts conditions.

The $S_j$ determine the weights, whereas $S_{\mathrm{FA}}$ estimates the risk after weighting. The term $-\mathrm{Disp}_f$ accounts for averaging predictions and $B$ for estimating their weights; the oracle bound replaces their signed sum by $\mathrm{Disp}_{\partial S}/T^2$. Its benchmark averages individual risks with nonrandom weights, not the risk of an optimal convex predictor. The bound holds for every fixed $T>0$ satisfying the conditions, and the remainder need not vanish on the local risk scale.

Polynomial growth of the maps and their first two partial derivatives is sufficient, not necessary, for \eqref{eq:gaussian-regularity}. The shrinkage maps below satisfy the required moments through bounded departures from the all-wide reference and bounded derivatives; Supplementary Sections S1.1 and S2 give the checks. Exactness concerns the smooth Gaussian model. Actual SMART fits retain treatment maximisation and estimate the representation parameters, so the remaining question is whether this Gaussian risk calculation evaluates the Q-function estimate in \eqref{eq:general-output}. Section~\ref{sec:implementation} establishes that connection at first order.

\section{Risk transfer and observable implementation}\label{sec:implementation}

To apply the Gaussian calculation to a SMART, we must construct its coordinates from observed regressions and retain the prediction error shared by the backward fits. We give conditions at a general target stage, then verify an observable two-stage implementation.

\subsection{Prediction-risk validity at a prespecified stage}
\label{sec:general-transfer}

Consider a $K$-stage SMART, with $K$ fixed as $n$ increases, and a target stage $t\in\{1,\ldots,K\}$ chosen before analysis. Write $Q_{t,n}^*$ for the target along a fixed local parameter sequence and $\mathcal H_t=L_2(\nu_t)$ for the prediction space. The key step is to express each fitted Q-function error as a common component plus the part represented by its recursive Gaussian map. The common component must remain in the full prediction risk even though it does not distinguish the fits.

Let $\bm\psi_t=(\psi_{t1},\ldots,\psi_{tm_t})^{\mathsf T}$ be fixed functions of target-stage history and treatment, representing the specification-dependent prediction-error directions. They are population basis functions, not re-estimated random functions in this representation. Choose them orthonormal under $\nu_t$, so $\E_{\nu_t}(\bm\psi_t\bm\psi_t^{\mathsf T})=I_{m_t}$ and $\Psi_t\bm v=\bm\psi_t^{\mathsf T}\bm v$ satisfies $\|\Psi_t\bm v\|_t=\|\bm v\|$. The vector $\vartheta_t$ collects the parameters determining the coordinate and prediction maps $f_{t,j}(\bm w;\vartheta_t)$ and $\bm A_t(\vartheta_t)$, such as regression cross-moments and covariance parameters. These are distinct from the local mean $\bm\delta_t$; in the two-stage model, $\vartheta_1=(\chi,\kappa)$. The vector $\bm w_t$ includes the required coordinates from all stages $t,\ldots,K$, and $\varsigma_t>0$ converts coordinate loss to prediction-risk units. Prespecified safeguards define fits and risk estimates on exceptional samples. The following conditions make the observable-coordinate approximation precise and ensure that its prediction-error expansion remains valid after adaptive weighting.

\begin{assumption}
For the fixed target $t$ and each fixed local parameter $\bm\delta_t$, suppose:
\begin{enumerate}
\item There are random elements $(U_{t,0,n},\bm w_{t,n}^0)$ on the probability space of the data used for fitting such that $(U_{t,0,n},\bm w_{t,n}^0)\Rightarrow(U_{t,0},\bm w_t)$ in $\mathcal H_t\times\R^{p_t}$, where $\bm w_t\sim N_{p_t}(\bm\delta_t,I_{p_t})$ and $\E\|U_{t,0}\|_t^2<\infty$. Observable estimates satisfy $\widehat{\bm w}_{t,n}-\bm w_{t,n}^0=o_p(1)$ and $\widehat\vartheta_{t,n}\to_p\vartheta_t$.
\item Jointly over the fixed finite set of Q-learning specifications,
\begin{equation}
 \sqrt n(\widehat Q_{t,n}^{(j)}-Q_{t,n}^*)
 =U_{t,0,n}+\varsigma_t\Psi_t
       \{f_{t,j}(\bm w_{t,n}^0;\vartheta_t)-\bm A_t\bm\delta_t\}
       +r_{t,j,n},
 \label{eq:general-representation}
\end{equation}
where $\max_j\|r_{t,j,n}\|_t=o_p(1)$, and the sequence
$n\max_j\|\widehat Q_{t,n}^{(j)}-Q_{t,n}^*\|_t^2$ is uniformly integrable.
\item For $\ell=1,\ldots,m_t$, the common component satisfies
\begin{equation}
 \E\{\langle U_{t,0},\psi_{t\ell}\rangle_t\mid \bm w_t\}=0.
 \label{eq:general-conditional-orthogonality}
\end{equation}
\item The target matrix $\bm A_t(\vartheta)$ is continuous at $\vartheta_t$, and the recursive maps and their first two partial derivatives with respect to $\bm w$ are jointly continuous in $(\bm w,\vartheta)$ near $\vartheta_t$. At the true representation parameters they satisfy Theorem 1, including $\E \mathrm{Disp}_{\partial S,t}<\infty$. The tuning parameter $T_t>0$, prior weights $\pi_{t,j}>0$ and library are prespecified and fixed.
\end{enumerate}
\end{assumption}

Condition (a) is verified by a joint central limit theorem for participant-level regression contributions and laws of large numbers for the estimated representation parameters. Condition (b) is obtained by propagating the regression errors in Proposition 1, retaining all local bias terms and controlling treatment switching; a uniformly bounded $(2+\epsilon)$ moment of the scaled prediction errors is sufficient for the required uniform integrability. These are statements about the data-generating model, not assumptions that risk already converges.

Condition (c) keeps the common error orthogonal after data-dependent weighting; orthogonality for each fixed fit alone is insufficient. In the regression construction, the oracle Bellman residual $\varepsilon_s^*=Y_s+V_{s+1}^*(H_{s+1})-Q_s^*(H_s,A_s)$ satisfies $\E(\varepsilon_s^*\mid H_s,A_s)=0$. Nested histories make the cross-stage covariances of the regression contributions $\phi_s(H_s,A_s)\varepsilon_s^*$ zero; within-stage residualisation also separates narrow and added blocks under homoscedasticity. In the two-stage construction below, the common component is a centred linear combination of these contributions, and their joint Gaussian limit verifies (c). These properties do not make nonlinear fitting errors independent. Supplementary Section S6.2 gives the regression argument. Condition (d) is checked from the recursive maps and their partial derivatives. Sections S3.5 and S4.10 give the general verification route and its two-stage verification, respectively.

For target $t$, evaluate \eqref{eq:fa-fixed-score}--\eqref{eq:fa-adapted-score} using $f_{t,j}$, $\bm A_t$, $T_t$ and $\pi_t$, and substitute $(\widehat{\bm w}_{t,n},\widehat\vartheta_{t,n})$ to obtain the weights $\widehat\alpha_{t,j,n}$ in \eqref{eq:general-output}. Denote the corresponding Gaussian risks by $R_{t,j}$ and $R_{\mathrm{FA},t}$, and the derivative dispersion by $\mathrm{Disp}_{\partial S,t}$. Theorem 2 shows that the Gaussian comparison then governs the full prediction risk of the returned Q-function, including its common error component.

\begin{theorem}
Under Assumption 1, for every prespecified $t\in\{1,\ldots,K\}$ satisfying its conditions, put $C_{0,t}=\E\|U_{t,0}\|_t^2$. Then
\begin{equation}
 \begin{split}
 n\cR_{t,n}^{(j)}&\longrightarrow C_{0,t}+\varsigma_t^2R_{t,j}(\bm\delta_t),\\
 n\cR_{t,n}^{\mathrm{FA}}&\longrightarrow C_{0,t}+\varsigma_t^2R_{\mathrm{FA},t}(\bm\delta_t).
 \end{split}
 \label{eq:general-risk-transfer}
\end{equation}
Moreover,
\begin{equation}
 \begin{split}
 \limsup_{n\to\infty}\Bigg[ n\cR_{t,n}^{\mathrm{FA}}
 &-\inf_{q\in\cW_J}\left\{
        \sum_jq_j n\cR_{t,n}^{(j)}
          +\varsigma_t^2T_t\KL(q\Vert\pi_t)\right\}\Bigg]\\
 &\leq\frac{\varsigma_t^2}{T_t^2}
              \E_{\bm\delta_t}\mathrm{Disp}_{\partial S,t}(\bm w_t).
 \end{split}
 \label{eq:general-oracle-transfer}
\end{equation}
If, additionally, $\widehat C_{0,t,n}\to_p C_{0,t}$ and $\widehat\varsigma_{t,n}\to_p\varsigma_t$, and the reported risk estimates
\begin{equation}
 \widehat S_{t,n}^{\mathrm{full}}
 =\widehat C_{0,t,n}+\widehat\varsigma_{t,n}^{\,2}
   S_{\mathrm{FA},t}(\widehat{\bm w}_{t,n};\widehat\vartheta_{t,n})
 \label{eq:general-reported-risk}
\end{equation}
are uniformly integrable, then
$\E\widehat S_{t,n}^{\mathrm{full}}-n\cR_{t,n}^{\mathrm{FA}}\to0$.
\end{theorem}

The proof is given in Supplementary Section S3.1.

The theorem completes the passage from coordinate risk to full Q-function prediction risk: the common term $C_{0,t}$ remains in each risk and cancels only in their comparison. The reported-risk conclusion is first-order unbiasedness, not consistent recovery of risk from one trial, and requires its own moment condition. When the conditions hold at every stage, each stage uses its own weights; the aggregates are not fed back into the separate backward fits. We next construct the observable quantities needed to apply this result.

\subsection{A two-stage model with an explicit recursive map}
To make the general result operational, we construct its coordinates and prediction map in a fully observed two-stage SMART with nonzero comparison feedback. This design is denoted E3$'$ in the accompanying computation files. Take independent $X_1,U_2\sim\mathrm{Unif}(-1,1)$ and independent balanced treatments $A_1,A_2\in\{-1,1\}$. Set $X_2=\rho X_1+(1-\rho)U_2+\omega A_1$, where $0<\rho<1$ and $\omega\geq0$. The rewards are
\begin{align}
 Y_1&=\bm\beta_1^{\mathsf T}(1,X_1,A_1,A_1X_1)^{\mathsf T}
                    +b_1X_1^2/\sqrt n+\varepsilon_1,\notag\\*
 Y_2&=\beta_{20}+\beta_{21}X_2+\beta_{22}A_1
       +A_2\{\tau_{20}+\tau_{21}X_2+b_2X_2^2/\sqrt n\}
       +\varepsilon_2.
 \label{eq:fa-dgp}
\end{align}
The Gaussian errors have variances $\sigma_s^2>0$ and are mutually independent and independent of the latent uniforms and randomised treatments. Both stages are observed for all participants. The terminal treatment coefficient is assumed uniformly positive for sufficiently large $n$.

The narrow vector at stage 1 is $n_1=(1,X_1,A_1,A_1X_1)^{\mathsf T}$, with added regressor $z_1=X_1^2$. At stage 2 it is $n_2=(1,X_2,A_1,A_2,A_2X_2)^{\mathsf T}$, with $z_2=A_2X_2^2$. Write $g_1=\operatorname{Var}(X_1^2)=4/45$ and $g_2=\operatorname{Var}(X_2^2)$. The limiting oracle stage-1 residual variance is
\begin{equation*}
 \bar\sigma_1^2=\sigma_1^2+(\beta_{21}+\tau_{21})^2(1-\rho)^2/3.
\end{equation*}
The local signals are $\delta_1=\sqrt{g_1}b_1/\bar\sigma_1$ and $\delta_2=\sqrt{g_2}b_2/\sigma_2$; write $\bm\delta=(\delta_1,\delta_2)^{\mathsf T}$. The feedback and transport coefficients are
\begin{equation}
 \chi=\frac{\rho^2\sigma_2\sqrt{g_1}}{\bar\sigma_1\sqrt{g_2}},\qquad
 \kappa=\frac{2\rho\omega\sigma_2}{\sqrt3\bar\sigma_1\sqrt{g_2}}.
 \label{eq:fa-geometry}
\end{equation}
Here $\chi\ne0$. The fixed continuation term $(\beta_{21}+\tau_{21})(\rho X_1+\omega A_1)$ lies in the stage-1 narrow span. For penalty pair $\lambda_j=(\lambda_{1j},\lambda_{2j})$, use $g_{s,j}=g_{\lambda_{sj}}$, where $g_\lambda(u)=u\{1+\exp(\lambda-u^2/2)\}^{-1}$; use the identity maps for the all-wide reference. Writing $\bm w=(w_1,w_2)^{\mathsf T}$, the Gaussian maps are
\begin{equation}
 f_j(\bm w)=\begin{pmatrix}
  g_{\lambda_{1j}}\{w_1+\chi g_{\lambda_{2j}}(w_2)\}\\
  \kappa g_{\lambda_{2j}}(w_2)
 \end{pmatrix},\qquad
 \bm A=\begin{pmatrix}1&\chi\\0&\kappa\end{pmatrix}.
 \label{eq:fa-e3-map}
\end{equation}
\Needspace{6\baselineskip}
The first coordinate targets $\delta_1+\chi\delta_2$; the second retains the narrow-span contribution needed for full initial-stage risk. With $q_j=w_1+\chi g_{\lambda_{2j}}(w_2)$, the feedback contribution to \eqref{eq:fa-fixed-score} is
\[
 2\chi^2
 \frac{\partial g_{\lambda_{1j}}(q_j)}{\partial q_j}
 \frac{\partial g_{\lambda_{2j}}(w_2)}{\partial w_2}.
\]
This term quantifies sensitivity propagated from the stage-2 fitted coefficient map into the stage-1 comparison. It vanishes when $\chi=0$ and is nonnegative for these increasing maps. The factor $\chi^2$ multiplies their local derivatives at the displayed inputs; because $q_j$ also changes with $\chi$, neither monotonicity in $|\chi|$ nor a reduction in prediction risk follows from this expression alone. Supplementary Section S3.3 gives the corresponding derivative-path calculation for general $K$. The remaining implementation task is to obtain the input coordinates and transport coefficients from the observed regressions.

\subsection{Computing the estimate from observed data}

Let $N_s$ be the sample narrow design, $r_s=(I-P_{N_s})z_s$ and
$\widehat g_s=r_s^{\mathsf T}r_s/n$. Superscripts $N$ and $W$
denote narrow and wide regression fits, respectively; when several
letters are used, they record the regression choices from the current
stage onward.

At stage 2, fit the wide model $\widehat Q_2^W$ and let
$\widehat c_2^W$ denote the fitted coefficient of its added regressor
$z_2$. Define
\[
 \widehat V_2^W(h_2)=\max_a\widehat Q_2^W(h_2,a).
\]
Next, fit the stage-1 wide model to the pseudo-outcome
$Y_1+\widehat V_2^W(H_2)$, and let $\widehat c_1^{WW}$ denote the
coefficient of its added regressor $z_1$. Thus the superscript $WW$
denotes the all-wide reference path: a stage-1 wide regression using
the continuation value generated by the stage-2 wide fit.

Let $\mathrm{RSS}_{2,W}$ and $\mathrm{RSS}_{1,WW}$ be the residual sums of squares from these reference regressions. Use the Gaussian working-likelihood residual scales
\[
 \widehat\sigma_2^2=\max\{\mathrm{RSS}_{2,W}/n,s_{\min}^2\},
 \qquad
 \widehat{\bar\sigma}_1^2=\max\{\mathrm{RSS}_{1,WW}/n,s_{\min}^2\},
\]
where the fixed positive floor $s_{\min}$ is below both limiting scales, and define
\[
 \widehat w_2
 =\frac{\sqrt{n\widehat g_2}\widehat c_2^W}{\widehat\sigma_2},
 \qquad
 \widehat\eta_1^{WW}
 =\frac{\sqrt{n\widehat g_1}\widehat c_1^{WW}}
        {\widehat{\bar\sigma}_1}.
\]

Evaluate the sample-residualised, standardised stage-2 added regressor
at the optimal action under the stage-2 wide reference fit, giving an
$n$-vector $v$. Residualisation is performed as a function of history
and action before evaluation; the supplement gives the explicit formula.
Set
\begin{align}
 \widehat\chi&=\frac{\widehat\sigma_2}{\widehat{\bar\sigma}_1}
                   \frac{r_1^{\mathsf T} v}{n\sqrt{\widehat g_1}},\qquad
 \widehat\kappa=\frac{\widehat\sigma_2}{\widehat{\bar\sigma}_1}
                   \sqrt{\|P_{N_1}v\|^2/n},\notag\\
 \widehat w_1&=\widehat\eta_1^{WW}-\widehat\chi\widehat w_2.
 \label{eq:fa-observable}
\end{align}
Write $\widehat{\bm w}=(\widehat w_1,\widehat w_2)^{\mathsf T}$. These coordinates and transport coefficients are computed from observed
regressions, without $\bm\delta$ or $Q^*$. On the common-action event
with inactive caps, the quadratic implementation has exact upstream
coordinate
$\widehat w_1+\widehat\chi g_{2,j}(\widehat w_2)$.
The subtraction in the definition of $\widehat w_1$ removes the
stage-2 wide-reference contribution from the stage-1 standardised
coefficient; each backward fit still uses its own nonlinear continuation.
Risk derivatives hold the estimated representation parameters fixed,
a replacement justified at first order. Supplementary Sections
S4.2--S4.3 give the sample identities. With these quantities available,
Algorithm 1 implements the combined estimate defined in
\eqref{eq:general-output} for $t=1$.

\Needspace{8\baselineskip}
\noindent\textbf{Algorithm 1. Feedback-aware tuning of completed Q-learning fits.}
\begin{enumerate}
 \item Fix the finite penalty-pair library, all-wide reference specification, tuning parameter, prior weights, regression safeguards and prespecified treatment tie-breaking rule. Compute the reference fits and \eqref{eq:fa-observable} from the observed data.
 \item Run Q-learning separately under each specification: combine its terminal narrow and wide fits, maximise its own fitted terminal value, and fit and combine its upstream regression models on that generated response.
 \item Evaluate the maps in the Gaussian shift model in \eqref{eq:fa-e3-map} and their risk estimates in \eqref{eq:fa-fixed-score} at $\widehat{\bm w}$ and the estimated transport coefficients. Compute the exponential weights in \eqref{eq:fa-weights}.
 \item Return the function
 \begin{equation}
  \widehat Q_1^{\mathrm{FA}}(h,a)
    =\sum_j\widehat\alpha_j\widehat Q_1^{(j)}(h,a).
  \label{eq:fa-output}
 \end{equation}
 Do not refit after this final combination. If reporting a risk estimate, include the outer-weight derivative correction and the common prediction component specified in the supplement.
\end{enumerate}

Two stagewise comparison criteria are supported. The quadratic version uses the squared Euclidean norm of the standardised coefficient vector, with common reference scales. The log-RSS version instead uses $n\log(\mathrm{RSS}_0/\mathrm{RSS}_1)$ from the regressions fitted under each specification, with weight $\{1+\exp(\lambda-\Lambda/2)\}^{-1}$. The pair $(1,1)$ is ordinary recursive Akaike weighting in the latter version and is both a library member and a separately reported comparator. The versions are first-order equivalent, not equal at finite $n$. The outer risk estimates use the same Gaussian approximation in both versions. Hard AIC is an external comparator, not a differentiable member of the tuning library.

\Needspace{8\baselineskip}
\subsection{Why the observable implementation inherits the guarantee}
Algorithm 1 uses estimated coordinates and actual treatment maximisation. Assumption 2 specifies the trial-law and safeguard conditions under which these empirical fits satisfy Assumption 1.

\begin{assumption}
In \eqref{eq:fa-dgp}, take fixed $b_1,b_2$ so that omitted added coefficients are of order $n^{-1/2}$. The terminal treatment coefficient is bounded away from zero and positive on the support for all sufficiently large $n$, and the population narrow and wide Gram matrices are positive definite. Use fixed positive empirical eigenvalue safeguards below the relevant population eigenvalues, a common zero-fit fallback on a required Gram failure, positive scale floors below the true scales, and fixed upper bounds for the transport coefficients strictly exceeding $|\chi|$ and $\kappa$, with nonnegative truncation for $\widehat\kappa$. The finite smooth library, tuning parameter $T>0$ and strictly positive prior weights are fixed. All safeguards and the treatment tie rule are prespecified. Unavailable reference coordinates and transport coefficients are set to zero on a Gram-failure sample for the asymptotic statement, while the common zero-fit prediction remains in the unconditional risk.
\end{assumption}

Bounded regressors, randomisation and continuous covariate variation give full-rank population Gram matrices. Gaussian errors and bounded transition noise give coefficient moment bounds; the separated terminal treatment gap yields $\rho_2(e_{2,n}^{(j)})=o_{L_2}(n^{-1/2})$ in prediction norm and after empirical stage-1 projection. Safeguards enclosing the population quantities are asymptotically inactive. Supplementary Section S4.10 provides the gap bound and verifies Assumption 1; absence of numerical failures alone does not verify these population conditions. The following corollary connects Algorithm 1 to Theorem 2.

\begin{corollary}
Under Assumption 2, the two-stage construction satisfies Assumption 1 at $t=1$, with $\varsigma_1=\bar\sigma_1$, $C_{0,1}=C_0$, and the maps in \eqref{eq:fa-e3-map}. In particular, $\widehat{\bm w}\Rightarrow N_2(\bm\delta,I_2)$ and $(\widehat\chi,\widehat\kappa)\to_p(\chi,\kappa)$. For the initial-stage SMART population evaluation law,
\begin{equation}
 \begin{split}
 n\cR_{1,n}^{\mathrm{FA}}&\longrightarrow C_0+\bar\sigma_1^2R_{\mathrm{FA}}(\bm\delta),\\
 n\cR_{1,n}^{(j)}&\longrightarrow C_0+\bar\sigma_1^2R_j(\bm\delta),
 \end{split}
 \label{eq:fa-smart-limit}
\end{equation}
with a common finite $C_0$. The conclusions of Theorem 2, including its oracle bound and first-order unbiased full-risk estimate, hold for both stagewise comparison criteria. The oracle remainder is $\bar\sigma_1^2\E_{\bm\delta}\mathrm{Disp}_{\partial S}/T^2$.
\end{corollary}

The verification and proof are given in Supplementary Section S4.

Corollary 1 establishes the full-risk guarantee for the Q-function actually returned by Algorithm 1, for both stagewise comparison criteria. The proof constructs $\widehat C_0$ and verifies that the centred common regression-error component remains orthogonal after weighting; no consistent estimate of $\bm\delta$ is required. The conclusion is pointwise along fixed local sequences. We now examine the accuracy of this risk assessment and the performance of the returned predictor at finite sample sizes.

\section{Numerical evaluation and a propagation diagnostic}
\label{sec:numerical}
The numerical study follows the two levels of the theory. We first check the exact risk identity in Theorem 1, then assess the finite-sample approximation and prediction performance of Algorithm 1. A separate simulated ADHD analysis illustrates the comparison-feedback mechanism of Section~\ref{sec:feedback}; it is not a validation of the tuning algorithm.

\subsection{Checking the Gaussian risk identity}
For the Gaussian calculations, we use the maps in \eqref{eq:fa-e3-map} with $(\rho,\omega)=(0.7,0.3)$ and $\sigma_2=\bar\sigma_1=1$, giving $(\chi,\kappa)=(0.437972,0.726967)$. Remaining reward coefficients are specified in Supplementary Section S5. The ten specifications are the nine pairs in $\{0.5,1,2\}^2$ and the all-wide reference, with uniform prior weights and prespecified $T=2$ for every signal. Supplementary sensitivity calculations at $T=1,2,4$ do not select $T$ from the data or claim that $T=2$ is optimal.

Gaussian quadrature evaluates prediction risk, its reported estimate and the oracle bound independently of finite-sample fitting. Table~\ref{tab:gaussian} compares direct risk with the expectation of \eqref{eq:fa-adapted-score}. The ``without $B$'' column omits only the outer-weight correction, retaining recursive feedback and dispersion subtraction. At $(\delta_1,\delta_2)=(1,1)$, it gives $1.444666$ instead of $1.696982$. Supplementary Section S5 reports the upper-risk expectation, oracle bound and the separate terms $\E B$, $\E\mathrm{Disp}_f$, $\E(B-\mathrm{Disp}_f)$ and $\E\mathrm{Disp}_{\partial S}/T^2$.

A separate ablation omits only the comparison-feedback derivative when choosing weights, keeping all backward fits unchanged. It agrees with FA at $\chi=0$ and can have either lower or higher risk at nonzero feedback. The supplementary results thus distinguish correctness of the risk criterion from universal superiority over an ablation.

\subsection{Prediction risk in finite SMART samples}
We next examine how accurately the Gaussian risk comparison predicts the finite-sample prediction-risk comparison between completed Q-learning fits. For $(\delta_1,\delta_2)\in\{(0,0),(0,1),(1,1),(2,3)\}$ and $n\in\{250,1000,4000\}$ we use $5{,}000$ independent trial datasets per cell. All methods are paired, and both stagewise criteria use the same $60{,}000$ datasets. The fitter receives only $(X_1,A_1,X_2,A_2,Y_1,Y_2)$ and estimates scales and transport coefficients. Supplementary Section S5 gives the signal-to-coefficient conversion, seed, safeguards, derivatives and evaluation Gram matrix.

Table~\ref{tab:fa-smart-gaps} shows the main log-RSS comparison. FA loses to recursive Akaike weighting at $(0,0)$ and $(0,1)$ and gains at $(1,1)$ and $(2,3)$. At $(2,3)$, hard AIC has lower limiting risk than both averaging methods (Table~\ref{tab:gaussian}). These comparisons support target-risk assessment, not a universal ordering of weighting rules.

The terminal treatment coefficient is positive over the bounded support, so population coefficient--Gram loss evaluates the true maximised $Q_1^*$. No safeguard or action-disagreement event occurred. Approximation error is nevertheless visible: at $(2,3)$ and $n=250$, full scaled risk is $11.6271$ (Monte Carlo standard error $0.1073$), whereas the mean first-order estimate is $11.0809$ ($0.0146$); the limit is $11.1799$. Supplementary tables report all risks and quadratic comparisons. These runs combine the effects of estimated scales, estimated transport coefficients and finite-sample coordinate distributions; they do not isolate variance estimation.

\subsection{Comparison feedback in a simulated ADHD trial}
The preceding experiments assess prediction risk; this illustration asks whether a changed model comparison must also change treatment recommendations. The public simulated \texttt{DTRlearn2::adhd} data contain 150 records designed to mimic a two-stage attention-deficit/hyperactivity-disorder SMART \cite{DTRlearn2,NahumShani2012}. In the source-aligned analysis, only the 99 nonresponders enter the second-stage regression; responders retain their observed outcome in the first-stage pseudo-outcome. This eligibility-gated illustration is not an application of the fully observed two-stage tuning theorem. Its narrower purpose is to display the motivating response-to-comparison pathway in a recognisable trial analysis.

The stage-2 candidate regression models vary first-treatment-by-second-treatment and adherence-by-second-treatment interactions. Under AIC, the two leading model weights are $0.245$ and $0.755$. Replacing hard downstream selection by averaging gives a first-stage pseudo-outcome root-mean-square difference of $0.0419$, gives a selection-minus-averaging difference of $-0.00808$ in the prior-medication-by-treatment coefficient of the same wide model, and changes its likelihood-ratio statistic by $0.240$. Yet the first-stage recommendations agree for all 150 records: the maximum contrast perturbation is $0.0121$, compared with a minimum absolute hard-selection contrast of $0.464$. The supplement provides the primary model definitions and diagnostic calculation. This illustrates why comparison feedback and crossing a treatment boundary are different events; it does not establish clinical benefit or validate the proposed fitter in a gated trial.

\section{Discussion}
\label{sec:discussion}

Feedback-aware soft tuning evaluates complete backward Q-learning fits at a common prediction target. Its risk criterion retains the effects of downstream fitting on upstream comparisons and separately accounts for estimating the final weights. The Gaussian risk identity, oracle inequality and transfer theorem connect this recursive dependence to observable Q-function prediction risk.

The analysis connects exact error propagation to observable risk assessment through residualised coefficients and differentiation of the full recursion and final weights. Risk transfer also requires a joint regression representation preserving orthogonality of the common prediction component after weighting. The theorem allows any prespecified target stage for fixed $K$; the two-stage construction verifies its conditions, and the supplement gives checks for other $K$-stage models.

The numerical results separate valid risk estimation from superiority over a comparator. Omitting the outer-weight correction understates risk in the reported Gaussian examples, while the sign of the Akaike-minus-FA gap changes across signals. Even where FA improves on recursive Akaike weighting, hard AIC can perform better. The finite-sample results show that a first-order risk estimate may remain optimistic in smaller trials. Sensitivity to prespecified $T$ also depends on the signal. These findings favour reporting full risks, paired differences and tuning sensitivity rather than selecting favourable settings after inspection.

In the simulated ADHD diagnostic, downstream averaging changes the upstream response, interaction coefficient and likelihood ratio without changing treatment recommendations because the perturbations are small relative to the fitted treatment margins. These quantities distinguish propagated model uncertainty from recommendation changes; the eligibility gate determines which participants transmit the perturbation.

Overall, this work provides a prediction-risk framework for complete backward Q-learning fits. It connects recursive model-comparison feedback, same-data aggregation and observable target-stage risk assessment. The general fixed-$K$ theorem requires an explicit joint representation and moment conditions; the two-stage construction verifies them rather than assuming that they hold for every SMART. The method thus evaluates model uncertainty jointly across stages while retaining the dependence created by backward recursion.

\section*{Data availability}
The diagnostic uses the public simulated \texttt{adhd} data in \texttt{DTRlearn2} 1.1 \cite{DTRlearn2}, not clinical records. Code, reference outputs and validation checks are archived on Zenodo (version 2.0.0) \cite{Kojima2026Code} (DOI: \href{https://doi.org/10.5281/zenodo.22700197}{10.5281/zenodo.22700197}); the development repository is \url{https://github.com/masahikoji/smart-qlearning-risk-propagation}.

\section*{Acknowledgements}
OpenAI's ChatGPT was used only to assist with manuscript proofreading and as an additional check of the simulation code and numerical results. The authors independently reviewed and verified the manuscript, code, numerical results, references, and scientific conclusions and take full responsibility for the work.

\noindent\textbf{Funding:} This work was supported by the Japan Society for the Promotion of Science KAKENHI [JP26K21185].

\begin{table}[!htbp]
\centering
\caption{Prediction risks and risk estimates under the Gaussian shift model}
\label{tab:gaussian}
\setlength{\tabcolsep}{5pt}
\begin{tabular}{@{}crrrrr@{}}
\toprule
$(\delta_1,\delta_2)$ & FA risk & $\E S_{\rm FA}$ & Without $B$ & Recursive Akaike & Hard AIC\\
\midrule
$(0,0)$ & 0.666478 & 0.666478 & 0.508100 & 0.629809 & 0.985891\\
$(0,1)$ & 1.035756 & 1.035756 & 0.841513 & 1.007531 & 1.527165\\
$(1,1)$ & 1.696982 & 1.696982 & 1.444666 & 1.710108 & 2.317610\\
$(2,3)$ & 2.440976 & 2.440976 & 2.394986 & 2.584688 & 2.160427\\
\bottomrule
\end{tabular}

\vspace{0.5em}
\begin{minipage}{0.96\linewidth}
\footnotesize
\textit{Note:} Results are for the two-stage feedback design and exclude the
common prediction component. Smaller values indicate lower prediction risk.
``Without $B$'' is
$\E(\sum_j\alpha_jS_j-\mathrm{Disp}_f)$ and therefore omits only the
outer-weight correction; it does not correspond to a differently fitted
aggregate.
\end{minipage}
\end{table}

\begin{table}[!htbp]
\centering
\caption{Finite-sample prediction-risk differences between recursive Akaike weighting and feedback-aware tuning}
\label{tab:fa-smart-gaps}
\setlength{\tabcolsep}{5pt}
\begin{tabular}{@{}crrrr@{}}
\toprule
$(\delta_1,\delta_2)$ & $n=250$ & $n=1000$ & $n=4000$ & Gaussian limit\\
\midrule
$(0,0)$ & \shortstack[r]{-0.0412\\(0.0020)}
        & \shortstack[r]{-0.0344\\(0.0017)}
        & \shortstack[r]{-0.0386\\(0.0019)}
        & -0.0367\\[0.5em]
$(0,1)$ & \shortstack[r]{-0.0292\\(0.0018)}
        & \shortstack[r]{-0.0309\\(0.0018)}
        & \shortstack[r]{-0.0288\\(0.0016)}
        & -0.0282\\[0.5em]
$(1,1)$ & \shortstack[r]{0.0111\\(0.0019)}
        & \shortstack[r]{0.0112\\(0.0018)}
        & \shortstack[r]{0.0111\\(0.0018)}
        & 0.0131\\[0.5em]
$(2,3)$ & \shortstack[r]{0.1623\\(0.0041)}
        & \shortstack[r]{0.1460\\(0.0036)}
        & \shortstack[r]{0.1403\\(0.0035)}
        & 0.1437\\
\bottomrule
\end{tabular}

\vspace{0.5em}
\begin{minipage}{0.96\linewidth}
\footnotesize
\textit{Note:} Entries are paired Monte Carlo estimates of
$n\{\cR_{1,n}^{\mathrm{Akaike}}-\cR_{1,n}^{\mathrm{FA}}\}$ from
$5{,}000$ feedback-design datasets per cell. Positive values favour
feedback-aware tuning. Parentheses contain Monte Carlo standard errors.
The separate Q-learning fits use log-RSS criteria, and the Gaussian limits
are obtained by independent quadrature.
\end{minipage}
\end{table}

\FloatBarrier
\bibliographystyle{unsrtnat}
\bibliography{main}

\end{document}